\documentclass[prl,twocolumn,10pt]{revtex4-2}
\usepackage{amssymb,amsmath}
\usepackage{graphicx}
\pdfoutput=1

\usepackage{placeins}
\usepackage{bm}
\usepackage{epstopdf}
\usepackage{color}
\usepackage{subfigure}
\usepackage{parskip}
\usepackage[compact]{titlesec}
\usepackage{hyperref}

\usepackage{esvect}

\definecolor{mycolor}{RGB}{10,10,10}
\hypersetup{
	colorlinks=true,
	linkcolor=black,
	filecolor=blue,      
	urlcolor=mycolor,
	citecolor = black,breaklinks,backref}

\usepackage{cancel}

\usepackage{placeins}

\titlespacing{\section}{0pt}{1em}{0em}
\titlespacing{\subsection}{0pt}{1em}{0em}

\begin{document}

\title{Topological directed amplification}
\author{Bikashkali Midya}
\email{midya@iiserbpr.ac.in} 
\affiliation{Department of Physical Sciences, Indian Institute of Science Education and Research Berhampur, India}

\begin{abstract}
A phenomenon of topological directed amplification of certain initial perturbations is revealed theoretically to emerge in a class of  asymptotically stable skin-effect  lattices described by nonnormal Toeplitz operators $H_g$ with positive ``numerical ordinate" $\omega(H_g)>0$. Nonnormal temporal evolution, even in the presence of global dissipation, is shown to manifest a counterintuitive transient phase of edge-state amplification---a behavior,  drastically different from the asymptote, that spectral analysis of $H_g$ fails to directly reveal. A consistent description of the effect is provided by the general tool of ``pseudospectrum", and a quantitative estimation of the maximum power amplification is provided by the {\it Kreiss constant}.  A recipe to determine an optimal initial condition that will attain maximum amplification power is given by singular value decomposition of the propagator  $e^{-i H_g t}$.   It is further predicted that the interplay between nonnormality and nonlinearity in a skin-effect laser array can facilitate narrow-emission spectra with scalable stable-output power. 
\end{abstract}
\maketitle

\section{I. Introduction}

Dynamical systems governed by Hamiltonians which fail to commute with their respective adjoints are called nonnormal in mathematical sense, and are in general extremely sensitive to boundary conditions. A class of such systems of enormous current interest is non-Hermitian topological lattices with nonreciprocal coupling \cite{Okuma2022a,Zhang2022,Bergholtz2021}. One of the exotic features of these systems is the manifestation of the non-Hermitian skin effect \cite{Lee2016,Yao2018,LeeThomale2019,Alvarez2018}, i.e., the localization of an arbitrary number of stationary states at one of the edges under open boundary conditions (OBCs); whereas bulk states remain extended in a closed lattices under periodic boundary conditions (PBCs). Origin of these localized skin modes is rooted in nontrivial topological winding of the bulk PBC spectral contour in complex-energy plane with respect to the interior OBC spectral points \cite{Gong2018,Zhang2020,Okuma2020,Wang2021,Borgnia2020}. The interplay of topology, non-Hermiticity, and non-normality is an active field of research both in theory \cite{Lee2019,JYLee2019,Kawabata2019,Longhi2022,Yokomizo2019,Herviou2019,Longhi2018b,Schomerus2020,Schomerus2021,Zirnstein2021,Haga2021,Okuma2021b,Teo2022,Longhi2020,Li2020,Longhi2019,Zezyulin2021,Okuma2021a,Xue2022,Li2020b,Lieu2018,Claes2021,McDonald2020,Scheibner2020,Ezawa2019,Guo2021,Song2019,Song2020,Longhi2015,Yoshida2020,Wojcik2020,Pyrialakos2022,Midya2018,Zhu2022,Wanjura2020,Ramos2021} and experiments \cite{Weidemann2020,Brandenbourger2019,Coulais2020,Gou2020,Lin2021,Zhang2021,Xiao2020,Helbig2020} with far reaching practical consequences.

An interesting question of fundamental importance, {\it how nonnormality influences the dynamical behavior of an injected power in a skin-effect lattice} remains largely unanswered. Here we uncover a peculiar phenomenon of topologically protected transient growth of initial energy directed towards an edge of a class of non-Hermitian photonic skin-effect lattices $H_g$. Such counterintuitive phase of directed amplification occurs,  although all modes decay monotonically, due to an interplay between topological skin-effect and non-orthogonality of the eigenbasis of a nonnormal Hamiltonian governing the dynamics. While dominant imaginary part of the corresponding spectrum correctly predicts asymptotic decay dynamics, the same fails to explain more complex transient amplification effect. A consistent description of the latter effect is provided by the mathematical notion of {pseudospectrum}, defined by the union of spectra of a Hamiltonian under exponentially small perturbations. Use of pseudospectrum to deal nonnormality has a rich tradition in linear algebra \cite{Trefethen,Bottcher}, its importance has been recognized in a variety of dynamical systems ranging from hydrodynamics \cite{Trefethen1993,Reddy1993}, and ecosystems \cite{Neubert1997}, optical waveguides \cite{Makris2014}, Lindblad master equation \cite{Okuma2022b}, nonnormal networks \cite{Asllani2018}, to  neuronal dynamics \cite{Murphy2009}.  Due to extreme sensitivity of a Hamiltonian with strong non-normality, a small perturbation can shift some of the pseudomodes into the upper half of the complex energy plane in favor of amplification. Remarkably, the amplification is topologically protected as long as corresponding pseudomodes, lying entirely inside the PBC spectral curve, are characterized by the underlying nontrivial topology of the system. Although a quantitative estimation of the maximum power amplification can be provided by the Kreiss matrix theorem \cite{Trefethen}, a critical question is {\it what initial perturbations give rise to the maximum transient growth}? A systematic procedure based on singular value decomposition (SVD) of the propagator $e^{-i H_g t}$ is explored here to reveal that an optimal initial condition that will attain maximum amplification is the principal right-singular vector. The corresponding transient time of amplification is shown to be approximated by the group velocity of the largest amplifying mode. A possibility of  single-mode emission with  enhanced output power is also discussed in a skin-effect laser array.

\begin{figure*}[htb!]
	\centering{
		\includegraphics[width=0.8\textwidth]{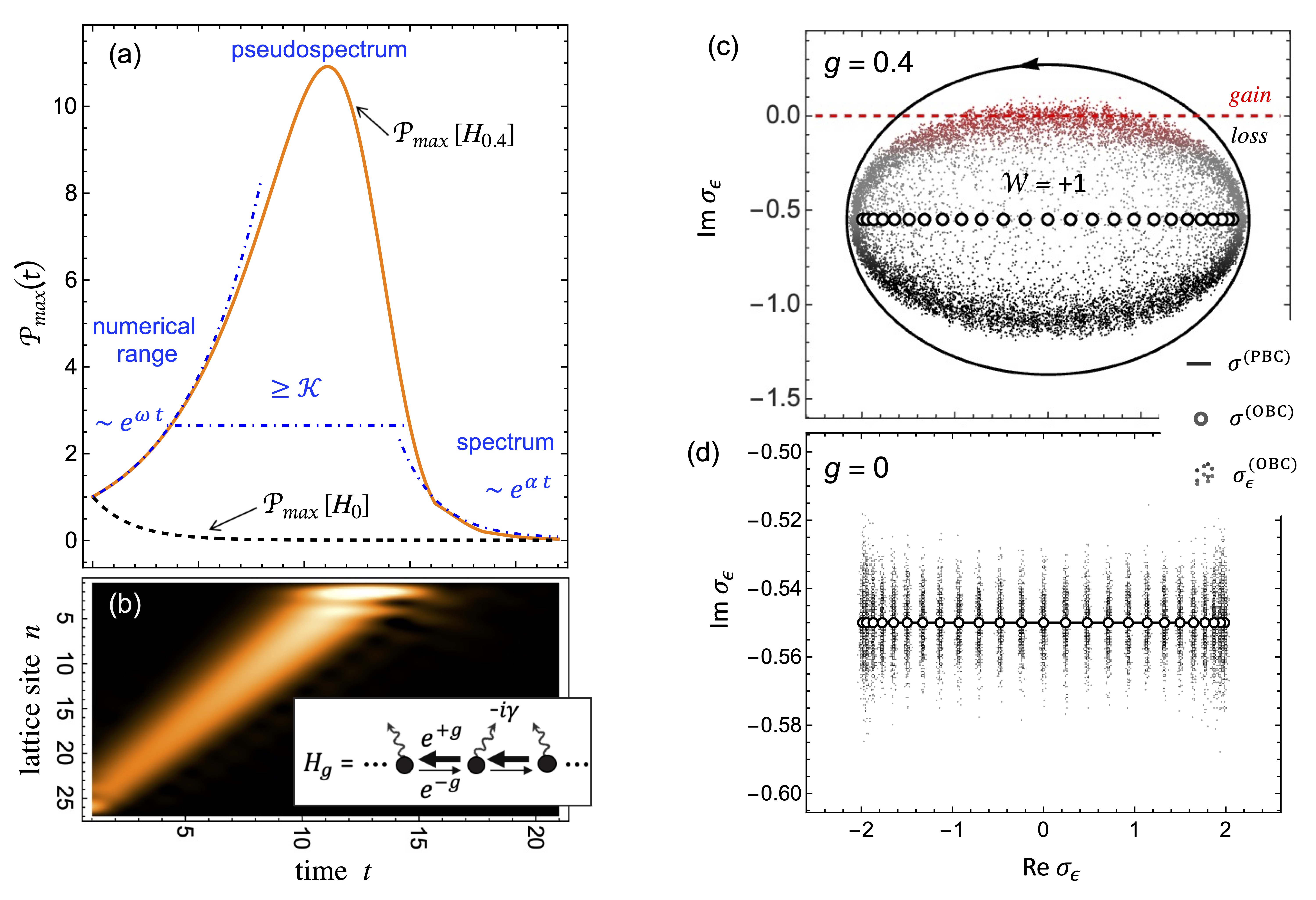} }
	\caption{(a) Magnitude of solutions of the dynamical Eq.~\eqref{Eq-3} yielding maximum  power is shown for a nonnormal lattice $H_{0.4}$ and a normal lattice $H_0$. The behavior of transient amplification of power can be described by the information of numerical range and pseudospectrum of the Hamiltonian, while the asymptotic decay is approximated by the spectrum.  A lower bound of the maximum power in the transient phase is given by the Kreiss constant $\mathcal{K}(H_{0.4})=2.6$. (b) The transient dynamics of intensity evolution of an initial condition [shown as `v' in Fig.~\ref{fig-3}(b)] which attains maximum power in the lattice $H_{0.4}$ showing monotonic growth of power directed to the boundary where skin-modes localize. (c) The OBC spectrum, PBC spectrum, and $\epsilon$-pseudospectrum of $H_{0.4}$ are shown for $\epsilon\approx0.15$.  Some pseudomodes shown in red dots in the gain region, $\operatorname{Im} \sigma_\epsilon>0$, originates from the center of unperturbed spectrum. (d) Pseudospectrum lies entirely in the lossy region, $\operatorname{Im} \sigma_\epsilon<0$, in the absence of nonnormality in $H_0$. Here $N=25$ and $\gamma=0.55$. All quantities are plotted in dimensionless units.} \label{fig-1}
\end{figure*}

\bigskip 

\section{II. Topological Hatano-Nelson lattice}
To illustrate the phenomena, we consider a generic nonnormal lattice given by the one-dimensional  Hatano-Nelson model \cite{Hatano1996}. The system is described by a family of Hamiltonians $H_g$ expressed in Toeplitz matrix form
\begin{equation}
{[H_g]}_{mn}=e^{-g}\delta_{m-1,n}+e^{g}\delta_{m+1,n}-i\gamma\delta_{m,n}\label{Eq-1}
\end{equation}
where $e^{-g}$, $e^{g}$, and $\gamma$ represent dimensionless nearest-neighbor rightward-, leftward-coupling amplitudes, and  onsite loss, respectively. Parameters $g$ and $\gamma$ are real, and taken to be non-negative for definiteness. The commutator $[H_g,H_g^\dag]=2\sinh 2g \operatorname{diag}(1,0,\cdots,0,-1)$ implies that the system is nonnormal for all $g\ne0$, while nonnormality is absent in $H_0$.  For a finite lattice with OBC, $H_g$  for $g\ne0$ and $g=0$ are nonunitarily equivalent by a gauge transformation $\vv{\Phi}_{g}=S\vv\Phi_{0}$, with $S_{mn}=e^{-gn}\delta_{m,n}$, and $\vv{\Phi}_g$ represents eigenvector of $H_g$.  As a consequence, the OBC spectrum of $H_{g\ne0}$ is spectrally isomorphic to that of $H_0$: $\sigma^{OBC}=\{2\cos~\ell \pi/(N+1) - i \gamma: \ell=1,2,\cdots N\}$, where $N$ is the number of sites in a lattice.
Components of stationary states $\vv{\Phi}_g(E_\ell\in~\sigma^{OBC})$ are given by $\phi_{n,\ell}=e^{-g n}\sin~\ell n\pi/(N+1)$, where $n$ is the lattice site index, showing the exponential localization, i.e. skin-effect, of all states at the left edge of a lattice $H_{g>0}$. However, skin-modes are not necessarily orthogonal as a consequence of the nonunitary transformation. This has an intriguing implication, as will be elaborated below, that the evolution dynamics of $H_g$ for $g\ne0$ is fundamentally different than that of $g=0$; a transient amplification phase exists in the former, while it never occurs in the latter case. Topological origin of the skin-modes are related to the nontrivial winding number of the PBC spectra characterized by \cite{Gong2018,Okuma2020}
\begin{equation}
	\mathcal{W}(z\in\sigma^{OBC})=\int_{0}^{2\pi} \frac{dk}{2\pi i}~\partial_k\operatorname{ln}\left[\sigma^{PBC}(k)-z\right], \label{Eq-2}
\end{equation}
where  the PBC spectrum for $g\ne0$ lies on an ellipse $\sigma^{PBC}~=~\{2\cos(k-ig)-i\gamma: k\in[0,2\pi]\}$, and collapses to $[-2-i\gamma,2-i\gamma]$ when $g=0$.
$\mathcal{W}=+1$ for $g>0$, and $0$ for $g=0$. The index theorem of a Toeplitz operator correlates analytical and topological index \cite{index}: $\operatorname{ind}H_g=~-\mathcal{W}$.

\bigskip

\section{III. Transient amplification}
The phenomenon of transient amplification is analyzed by considering how an initial perturbation $\vv{\Psi}_g(t=0)$ evolves in a dissipative Hatano-Nelson lattice by the dynamical equation of motion 
\begin{equation}
	i\dot{\vv\Psi}_g(t)=H_g \vv{\Psi}_g(t), ~~~ ||\vv{\Psi}_g(t=0)||=1.\label{Eq-3}
\end{equation}
 The general solution of Eq.~\eqref{Eq-3} is given by $\vv{\Psi}_g(t)=G(t)\vv{\Psi}_g(t=0)$, where $G(t)=e^{-iH_gt}$ is the propagator.  
A physical observable of interest is the optical power determined by the Euclidean norm of the solution  $\mathcal{P}(t)=||\vv{\Psi}_g(t)||=\left(\sum_n|\psi_n(t)|^2\right)^{1/2}$ (square root of the actual power is considered for the convenient description of what follows henceforth), where $\psi_n$'s are modal amplitudes at site $n$. 

{\it Amplification in a Hatano-Nelson Dimer}.
As an example, we first consider a dimer amenable to analytical treatment. The eigenvalues and corresponding normalized eigenvectors of the dimer Hamiltonian 
\begin{equation}
	H_g=
	\begin{bmatrix}
		-i\gamma & e^{g}  \\
		e^{-g} & -i\gamma 
	\end{bmatrix},
\end{equation}
are given by
\begin{equation}
	E_\pm=\pm 1-i\gamma, ~~~ \vv{\Phi}_{g,\pm}= \frac{1}{\sqrt{e^{2g}+1}}\begin{bmatrix}
		 e^{g}   \\
		\pm 1
	\end{bmatrix}.
\end{equation}
Note that the above eigenvectors are not orthogonal (with an exception for $g=0$) and approach each other in the large $g$ limit, as is evident from the angle between two eigenvectors 
\begin{equation}
	\theta = \cos^{-1}(\tanh g) \rightarrow 0  ~~~\mbox{as} ~~g\rightarrow\infty.
\end{equation}
Now consider time evolution of a specific initial state $\vv\Psi_g(t=0)=\left[-i/\sqrt{2},1/\sqrt{2}\right]^T$.  The solution at a later time is given by
 \begin{equation}
 	\vv\Psi_g(t)\propto(e^{g}-i)e^{-iE_+t}\vv\Phi_{g,+} -(e^g+i)e^{-iE_-t}\vv\Phi_{g,-}
 	\end{equation} 
apart from a multiplicative constant $2^{-3/2}(e^{-2g}+1)^{1/2}$. Even though the magnitude of solution eventually decays in time, it can initially grow due to non-orthogonal superposition of two eigenvectors that decay at different rates as time evolves. This is clear from corresponding time-dependent power 
\begin{equation}
\begin{matrix}	\mathcal{P}(t)=e^{-\gamma t}(\cos^2t+ \cosh 2g \sin^2t+\sinh g\sin 2t)^{\frac{1}{2}}  \\\\
	\simeq 1+(\sinh g -\gamma)t +\mathcal{O}(t^2) ~~~\mbox{as}~~~ t\rightarrow 0,
	\end{matrix}
\end{equation}
that decays for large time, but experiences initial amplification provided $\gamma<\sinh g$ and $g>0$. 

{\it Amplification in a Hatano-Nelson lattice}: For a larger lattice, we first investigate maximum power that can be attained. Using the inequality  $||AB||\le ||A||~||B||$ and the general solution of Eq.~\eqref{Eq-3}, we obtain the maximum power:
\begin{equation}
 \mathcal{P}_{\max}(t)= \sup_{||\vv{\Psi}_g(t=0)||=1} ||G(t)\vv{\Psi}_g(t=0)||=||G(t)||, \label{Eq-4}
	\end{equation}
given by the $2$-norm, i.e. maximum singular value, of $G(t)$ at each time $t$. Note that $\mathcal{P}_{\max}(t)$ is different for different Hamiltonian $H_g$. For a given $H_g$,  $\mathcal{P}_{\max}[H_g](t)$ represents an envelop of all possible power curves corresponding to the evolution of different initial conditions under $H_g$. Two particular examples of $\mathcal{P}_{\max}$ are presented  in Fig~\ref{fig-1}(a) corresponding to a nonnormal lattice $(g=0.4)$ and a normal lattice $(g=0)$, showing fundamentally different behavior. A remarkable effect of the short-time amplification phase, even in the presence of loss, is observed in the nonnormal lattice before the power decays asymptotically to zero. An intensity evolution pattern in $H_{0.4}$ corresponding to an initial excitation at the right edge favoring maximum amplification is shown in Fig.~\ref{fig-1}(b); intensity evolves unidirectionally and power grows monotonically.  The power becomes maximal when the excitation reaches at the left edge of the lattice where skin-modes are localized. Note, however, that all skin-modes in general attenuate in a lattice with non-zero $\gamma$. Corresponding spectrum of the lattice, therefore, provides no information about the observed amplification. It is the emergence of pseudomodes (or quasiedge modes \cite{Gong2018}) in a transient time is responsible for the observed amplification. A complete description of this  effect is given below by the notions of ``numerical range" and ``pseudospectrum", well known in matrix analysis \cite{Trefethen}.

{\it Asymptotic phase, $t\rightarrow\infty$}: 
In this regime, the general solution of Eq.~\eqref{Eq-3} reduces to $\vv\Psi_g(t\rightarrow\infty)~\sim~ e^{\alpha(H_g)t}$, where $\alpha(H_g)$ is the `spectral ordinate' \cite{note} given by $\alpha~=~\sup~\operatorname{Im}~\sigma^{OBC}= -\gamma$, independent of the parameter $g$. This implies that solution and hence power decays to zero asymptotically whenever $\gamma\ne0$, $\forall g$.

{\it Initial phase, $t\rightarrow 0$}:  According to the Hille-Yoshida theorem \cite{Trefethen}, the behavior of $\mathcal{P}_{\max}$ is given by
\begin{equation}
	\lim_{t\rightarrow0}~ \frac{d}{dt} \mathcal{P}_{\max}[H_g](t) = \omega(H_g),\label{Eq-5}
\end{equation}
where $ \omega(H_g)$ is the ``numerical ordinate" \cite{note} defined by the maximum among imaginary parts of numerical range of $H_g$ \cite{Neubert1997,Asllani2018,Trefethen}, and determined analytically by
 \begin{equation}
\hspace{-.15cm} \omega=\sup \sigma^{OBC} ~\frac{H_g-H_g^\dag}{2i}=2 \sinh |g| \cos\frac{\pi}{N+1}-\gamma.\label{Eq-6}
\end{equation}
According to Eq.~\eqref{Eq-5}, $\omega(H_g)$ is the slope of the curve $\approx e^{\omega(H_g) t}$ which approximately fits $\mathcal{P}_{\max}(t\rightarrow0)$. The result obtained in Eq.~\eqref{Eq-6} is, therefore, key to predict the transient growth at the outset; $\omega>0$ implies the onset of energy growth with growth rate is given by $\omega$, and a negative value of $\omega$ implies an opposite behavior. For $H_0$, $\omega(H_0)=-\gamma<0$ implies that  amplification is not possible in a dissipative lattice in the absence of non-normality. Conversely, for a given $\gamma\ne0$, it is always possible to find critical parameters $(g,N)$ for which $\omega(H_g)>0$, indicating the possibility of amplification in a nonnormal lattice $H_{g\ne0}$.  For the example in Fig.\ref{fig-1}(a), $\omega(H_{0.4})=0.27$. 

In an amplifying phase, Eq.~\eqref{Eq-6} further indicates that the larger the values of $g, N$, and hence $\omega$, the steeper the initial amplification rate. This is a direct consequence of the fact that deviation of $H_g$ from the normality, measured by the Henrici's departure from normality \cite{Asllani2018,Trefethen}: $\operatorname{d}(H_g)~=~[\operatorname{tr}({H_{g}}^\dag H_g- {H_{0}}^\dag H_0)]^{1/2}=2\sqrt{N-1}\sinh g$, increases with either the nonreciprocal coupling parameter $g$, or/and the geometric dimension $N$ of the lattice.

{\it Intermediate phase, $t\rightarrow$ finite}: Having obtained the initial trend for the maximum power curve, we now describe the actual physics behind the amplification. Note that a non-normal system is highly sensitive to perturbation. A convenient tool to investigate perturbed system is the method of pseudospectrum. For $\epsilon>0$, the $\epsilon$-pseudospectrum of $H_g$ under OBC is defined by \cite{Trefethen} $\sigma_\epsilon^{OBC} =\{z\in \mathcal {C} : ||(z-H_g)^{-1}||>\epsilon^{-1}\}$. For efficient computation of pseudospectrum  we consider collection of spectra of a perturbed Hamiltonian such that 
\begin{equation}
	\sigma_\epsilon^{OBC}(H_g)=\{z\in \mathcal {C} : z\in \sigma(H_g+H'), ||H'||\le \epsilon\}.
\end{equation} 
It has the properties that $\sigma^{OBC}\subseteq\sigma_\epsilon^{OBC}$, $\lim_{\epsilon\rightarrow0}\sigma_\epsilon^{OBC}~=~\sigma^{OBC}$, and $\lim_{\epsilon\rightarrow0,N\rightarrow\infty}\sigma_\epsilon^{OBC}=\sigma^{PBC}$. Contrary to the normal system $H_0$, which is less sensitive to the perturbations and corresponding  $\epsilon$-pseudospectrum lies within the $\epsilon$-neighborhood of the spectrum, perturbation to a nonnormal system $H_{g\ne0}$ yields drastic change in the spectrum. For $H_g$ with sufficiently strong nonnormality, the corresponding unperturbed spectrum initially belongs to negative imaginary plane can protrude to the positive imaginary plane under small perturbation. This has the remarkable consequence that the dynamical evolution which, according to spectrum, must decay, but actually can grow upon perturbation by the pseudomodes with positive imaginary pseudo-eigenvalues. Refer to Fig.~\ref{fig-1}(c) for a particular example of pseudospectrum corresponding to the amplification shown in Fig.~\ref{fig-1}(a). 

Two important observations are as follows. First, pseudo edge-modes which belong to the center of unperturbed OBC spectrum become amplifying first.  These modes have largest group velocity (discussed in Sec.~V) and are perturbed most. On the other hand,  because of the negligibly small group velocity, modes at the edge of the unperturbed spectrum are least affected by the perturbation.  Second, $\epsilon$-pseudospectrum resides entirely inside the PBC spectrum for small enough $\epsilon$. According to Eq.~\eqref{Eq-2}, all the pseudomodes, therefore, satisfy $\mathcal{W}(z\in\sigma_{\epsilon}^{OBC})=+1$. This implies that the effect of nonnormality induced transient growth is topologically protected. 

\begin{figure}[t!]
	\centering
	\includegraphics[width=0.49\textwidth]{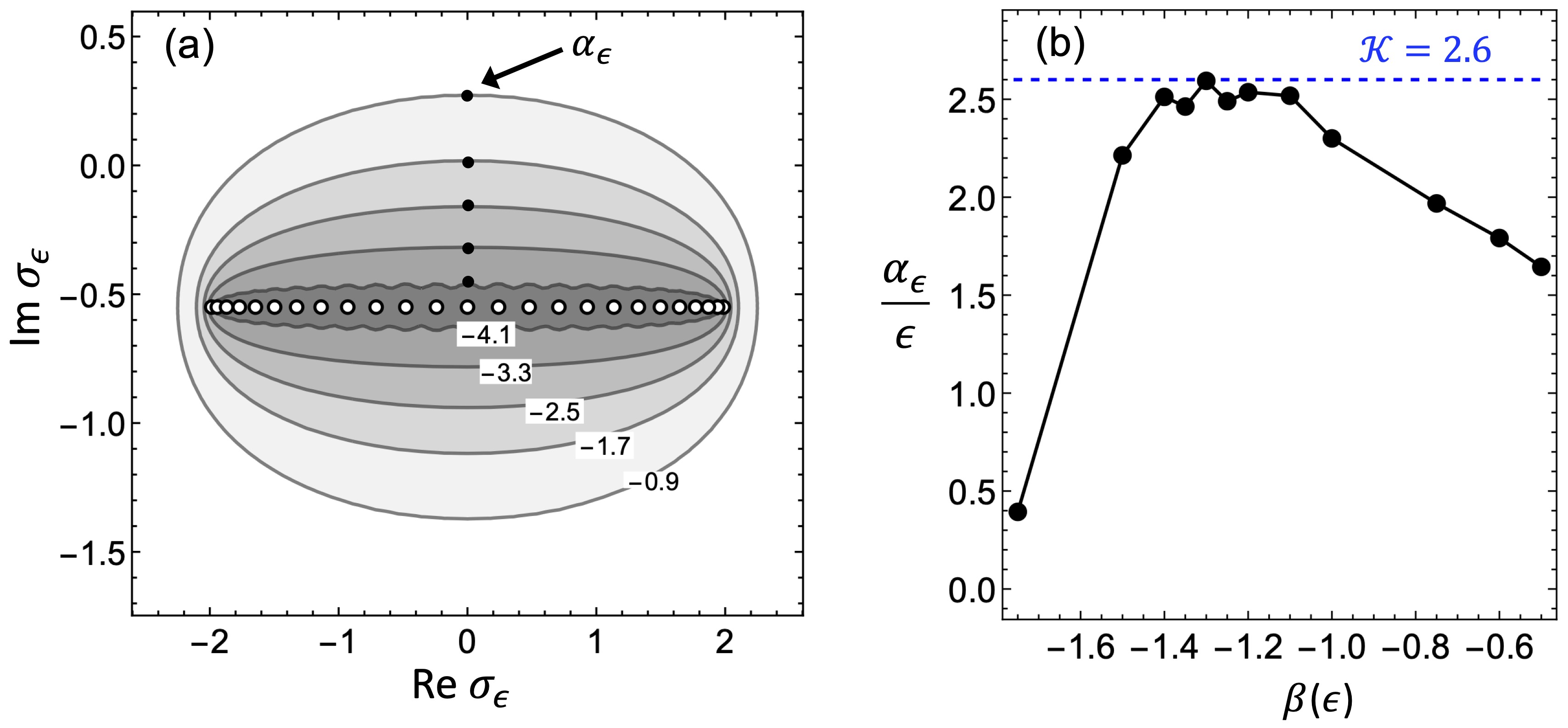} 
	\caption{(a) Pseudospectral ordinates $\alpha_\epsilon(H_{0.4})$ are determined from the level curves $||\left(z-H_{0.4}\right)^{-1}||=1/\epsilon$ shown here for different small perturbations $\epsilon=10^{\beta}$, number on each contour represents a value of $\beta$. (b) Kreiss constant $\mathcal{K}(H_{0.4}) \simeq 2.6$ is the maximum of $\alpha_\epsilon/\epsilon$ $\forall\epsilon\ll1$.} \label{fig-2}
\end{figure}

The size of a maximum transient growth can be quantitatively estimated by the Kreiss matrix theorem~\cite{Trefethen}. In particular, $\mathcal{K}(H_g)\le\sup_t\mathcal{P}_{\max}[H_g](t)\le e N \mathcal{K}(H_g)$, where the {\it Kreiss constant} $\mathcal{K}$ is defined in terms of $\epsilon$-pseudospectral ordinate $\alpha_\epsilon$ such that
\begin{equation}
	\mathcal{K}(H_g)=\sup_{\epsilon>0} ~{\alpha_\epsilon(H_g)}/{\epsilon}.
\end{equation}
$\mathcal{K}>1$ indicates that there must be amplification in the system. For the example in Fig~\ref{fig-1}(a), $\mathcal{K}(H_{0.4})=2.6$  obtained from the pseudospectral level curves [see Fig.~\ref{fig-2}].  

\begin{figure}[t!]
	\centering
	\includegraphics[width=0.35\textwidth]{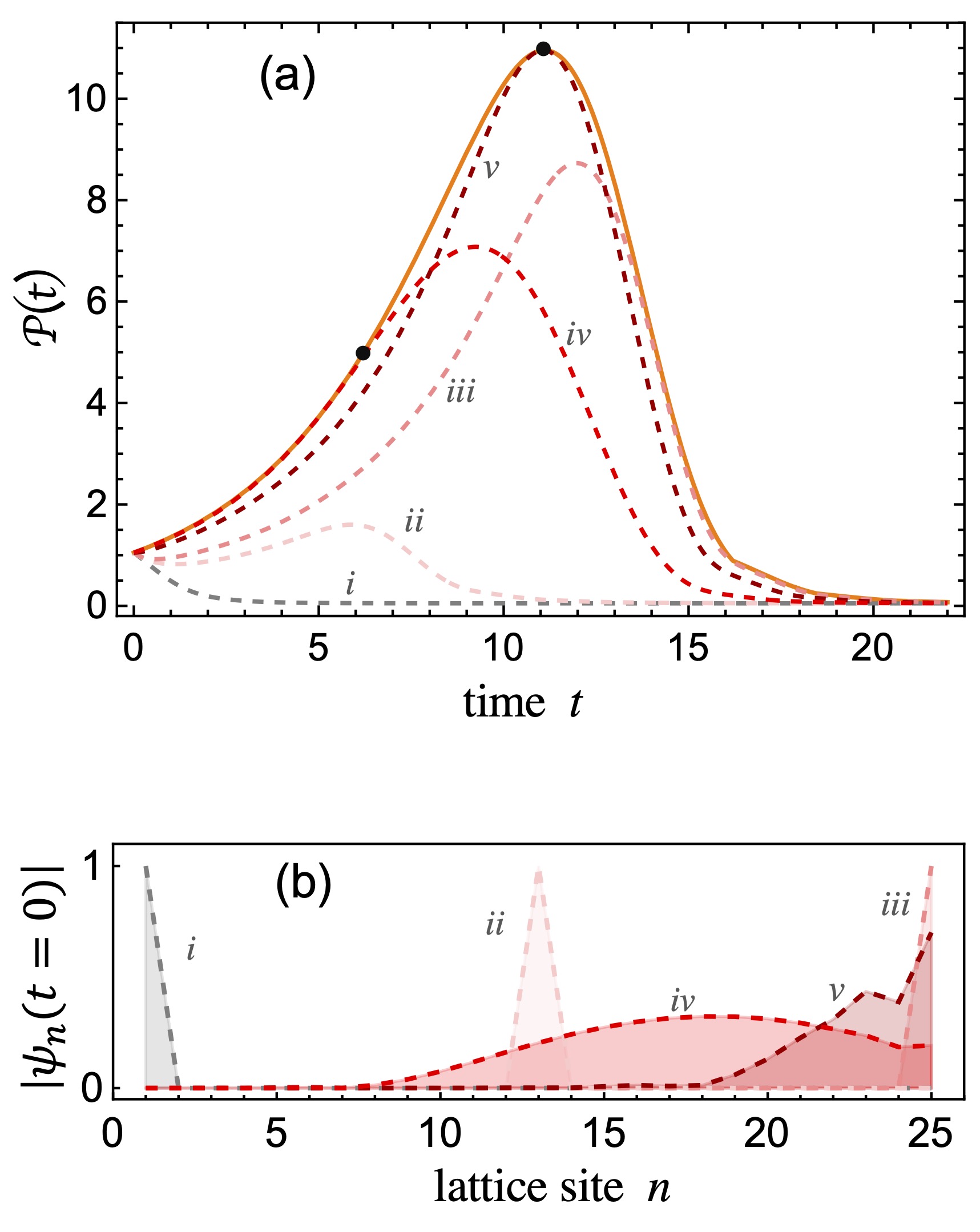} 
	\caption{(a) Transient dynamics of different initial conditions shown in (b) for a lattice $H_{0.4}$. `i', `ii', and `iii' in (b) are single site excitations, while `iv' and `v' are obtained by SVD of the propagator $e^{-iH_{0.4}t}$ at time $t=6$, and $t=11$, respectively. Solid line in (a) representing the envelop of $\mathcal{P}_{\max}[H_{0.4}]$. Optimal initial perturbation, marked as `v' and `iv', touches the envelop curve acquire maximum power amplifications shown by black dots at $t=11$ and $t=6$, respectively.  }\label{fig-3}
\end{figure} 

\bigskip 

\section{IV. Optimal initial condition for maximum amplification}
Above analysis of $\mathcal{P}_{\max}$ indicates that the transient growth of energy is possible in a nonnormal system. In practice, however, the size of amplification depends on particular form of an initial condition. Optimal initial perturbation that will reach maximum amplification $\mathcal{P}_{\max}(t)$ at time $t$ can be obtained by the method of singular value decomposition (SVD) of the propagator: $G=U\Sigma V^\dag$, where diagonal entries $\Sigma_{nn}$ are singular values, and corresponding left- and right-singular vectors are given by the columns of the matrices $U$ and $V$, respectively. If $\Sigma_{11}$ denotes the largest singular value of $G(t)$, then SVD $G\vv{v}_1=\Sigma_{11} \vv{u}_1$ describes that $G$ maps an initial vector $\vv{v}_1$ to an output vector $\vv{u}_1$ amplified by $\Sigma_{11}=\mathcal{P}_{\max}(t)$ [by Eq.\eqref{Eq-4}].  It is therefore appropriate to consider $\vv{v}_1$ as the initial condition for the Eq.~\eqref{Eq-3} to achieve maximum amplification power at time $t$.

The general idea above is exemplified in Fig.~\ref{fig-3}  for $H_g$ with $g=0.4$. The initial conditions [marked  ``iv" and ``v" in the Fig.~\ref{fig-3}(b)] obtained by the SVD of $G(t)$ at time $t=6$ and $t=11$ respectively, yields maximum amplification at respective times. For comparison, single site excitation at three different locations of the lattice are also considered as initial conditions. It is seen that single site excitation away from the skin-mode localization center provides in general pretty good amplification. However, the excitation one at the left edge [marked  ``i" in Fig.~\ref{fig-3}(b)] does not amplify. Strictly speaking, {\it non-normality of a lattice is not sufficient for observing transient amplification}, shape of initial profile is crucially important. 

\bigskip

\section{V. Group Velocity and  transient time}
Here, we discuss how to estimate transient time $\tau$ at which maximum amplification primarily occurs at the edge where skin-modes are localized. $\tau$ is determined by the time elapses when a localized wave-packet propagates unidirectionally to the skin-edge starting from the site of initial excitation, usually at the opposite edge, thus traversing a distance proportional to the lattice length $(N-1)$. Transient time has previously shown to be related to Lieb-Robinson bound and time-energy uncertainty relation \cite{Gong2018,Okuma2022b}. Here we follow an alternative route and show that the same can be computed by the knowledge of group velocity of Floquet-Bloch modes. Group velocity of a wave packet with wave number $k$ is obtained from the energy-momentum relation under PBC \cite{Longhi2015}:
\begin{equation}
	v(k,g)={\rm Re}\left(\frac{d\sigma^{PBC}}{dk}\right)=-2\cosh g ~\sin k.
\end{equation}
It is worth mentioning the implications of the group velocity. We first consider $g>0$ and $\gamma=0$. In this case, waves with $0<k<\pi$ (belong to upper semi-ellipse of PBC spectra) have negative group velocity as well as Im~$\sigma^{PBC}>0$. These waves therefore propagate to the leftward direction and are amplified. On the other hand, waves with $\pi<k<2\pi$ (belong to lower semi-ellipse) have positive group velocity and Im~$\sigma^{PBC}<0$, implying that rightward propagating waves are attenuated and eventually not observable. This asymmetry between left- and rightward propagating waves explains the unidirectional energy flow in a skin-effect lattice. In the presence of loss ($\gamma\ne0$ and $\gamma<2\sinh g$), however, leftward amplifying waves belong to much narrower interval centered at  $k\sim\pi/2$, and the interval for attenuating modes (which now includes all rightward, and some leftward waves as well) broadens compared with $\gamma=0$ case, as exemplified in Fig.~\ref{fig-1}(c). Thus unidirectional propagation persists in the presence of moderate nonzero loss [Fig.\ref{fig-1}(b)].  

To determine the time of maximum amplification by leftward propagating pseudomodes, we note that wave with $k=\pi/2$ (which belongs to center of the unperturbed spectrum) has the largest group velocity i.e. $|v(\pi/2,g)|=2 \cosh |g|=v_{\rm{max}}$. This has physical implication that the corresponding pseudomodes lying at the center of the spectrum amplifies most  under perturbation [Fig.\ref{fig-1}(c)].  (Note that group velocity is maximum for another central mode at $k=3\pi/2$, but this mode suffers maximum loss and is not relevant in transient amplification.) The transient time therefore approximately given by 
\begin{equation}
	\tau\simeq \frac{N-1}{v_{\rm{max}}}=\frac{N-1}{2 \cosh |g|}\label{Eq-15}
\end{equation}
For the example in Fig.~\ref{fig-1}(b), $\tau\simeq11$. For a bulk excitation, however, traveling distance and hence the value of $\tau$ reduces compared to an edge excitation [e.g. ``ii" and ``iv" in Fig.~\ref{fig-3}].  Equation \eqref{Eq-15} implies that transient time can be exponentially lowered  by increasing left or right hopping imbalance parameter, i.e $|g|$. 

\bigskip

\section{VI. Nonnormal laser array}
\begin{figure}[t!]
	\centering
	\includegraphics[width=0.49\textwidth]{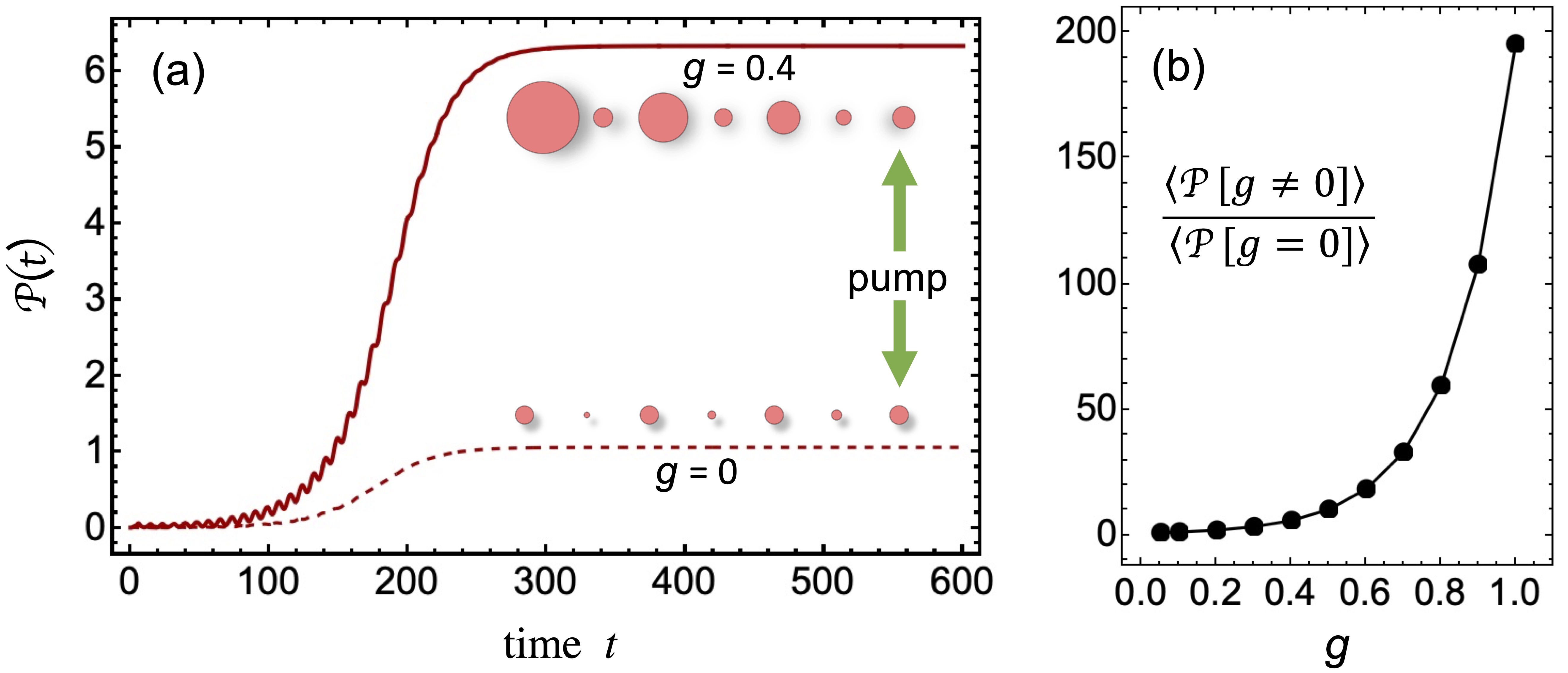} 
	\caption{ (a) Emission powers of an array consisting $N=7$ coupled lasers are shown for a nonnormal system $g=0.4$ and a normal lattice $g=0$. In both cases, $\gamma=0.1$ and single site at the right edge is pumped with $\Gamma_n=0.5\delta_{n,7}$. Absence of oscillation in power curve after an initial transient period implies that the system is dynamically stable. Relative intensity distribution in individual lasers at time t=1000 are shown by filled discs. (b) Ratio of time-averaged emission powers between a nonnormal and normal lattices vs the coupling parameter $g$ shows the exponential increment of power in the nonnormal array.  }\label{fig-4}
\end{figure}

Finally, we briefly report the interplay between nonnormality and nonlinearity in a nonreciprocally coupled microring laser array. Laser dynamics is determined by $i\dot{\vv{\Psi}}~=(H_g+H_{NL})\vv{\Psi}$, where
\begin{equation}
[H_{NL}]_{mn} = \frac{i\Gamma_n}{1+|\psi_n|^2} \delta_{m,n} 
\end{equation}
with $\Gamma_n$ represents site-dependent saturable gain \cite{Harari2018}. In the absence of gain, passive skin-modes are localized at the left edge of a lattice with $g>0$ and only clockwise circulation in each of the rings is assumed. When gain is switched on by pumping only a single laser at the right edge $\Gamma_n=\Gamma\delta_{n,N}$ [corresponds to the initial condition `iii' for a linear lattice, refer to Fig.~\ref{fig-3}(b)], lasing emission mainly occurs from the left-edge of the array [Fig.~\ref{fig-4}(a)]. This implies that linear skin-effect sustains under nonlinearity. The `nonlocal' pumping strategy i.e. away from the lasing output, not only convenient for practical implementation in experiment, but also is numerically verified to provide temporally stable emission (which is challenging to achieve in a large-scale laser array \cite{Longhi2018a}). Total output power is seen to scale exponentially, compared with a conventional array with reciprocally coupled lasers, when degree of nonnormality is increased by tuning the nonreciprocal coupling parameter $g$ (elaborated in Fig.~\ref{fig-4} with example). 

 A natural question arises here regarding the role of linear transient amplification in the lasing behavior. Note that linear amplification is not necessary for lasing to occur under nonlinear gain. However, maximally growing linear pseudomode (emerging from the center of unperturbed spectrum) in the transient period possesses least lasing threshold and thus favorable for lasing action when nonlinear gain saturation sets in.  Linear amplification therefore leverages reduction of nonlinear mode competition and hence helps narrow-spectral emission necessary for laser stability \cite{Midya2019}.  This principle remains intact in a higher-dimensional skin-effect laser array, and qualitatively explains anomalous single-mode lasing previously predicted in Refs.~\cite{Teo2022,Zhu2022}.

\bigskip

\section{VII. Conclusion and Outlook}
It is demonstrated that nonnormal nature of a class of recently-discovered non-Hermitian skin-effect lattices can give rise to a fundamentally new effect of topologically-protected one-way transient amplification of injected energy.  Condition of amplification is analytically derived, bounds of maximum amplification, and optimal initial condition to attain it are determined by the pseudospectrum of the Hamiltonian, and SVD of the propagator, respectively. Time of transient amplification is shown to be approximated by the maximum group velocity of Floquet-Bloch waves. Potential of exponential enhancement of stable-emission power is revealed in a nonnormal laser array. Reported results can be extended to higher-dimensional lattice models, and can be tested in quantum and classical topological devices with engineered nonnormality, particularly for directed transport or sensing of weak signals, and large-scale high-power laser applications. 

Few important remarks are in order here.
 
 A reciprocal lattice with passive parity-time (PT) symmetry 
 	\begin{equation*}
 		{[H_{PT}]}_{mn}=\kappa\delta_{m-1,n}+\kappa\delta_{m+1,n}-i\gamma[1+(-1)^n]\delta_{m,n},
 	\end{equation*}
 	 is a special class of asymptotically stable nonnormal system where skin-effect as well as dynamical transient-amplification are absent in both broken and unbroken regimes. Absence of the latter effect can be justified by vanishing numerical ordinate $\omega(H_{PT})=\sup\{0,-2\gamma\}=0$ (see Eq.~\eqref{Eq-6} and discussion therein) for all values of system parameters.  Maximum power corresponding to the Hamiltonian defined above is bounded below and above by $e^{\alpha(H_{PT}) t}\le\mathcal{P}_{max}\le e^{\omega(H_{PT})t}=1$, where $\alpha(H_{PT})<0$.  Nonnormal lattices with skin-effect thus represent a novel class of materials with fundamentally different amplification property absent in PT-symmetric systems.

Skin-effect assisted amplification has previously been pointed out for noninteracting electrons \cite{Gong2018,Okuma2022b}. Photonic lattices, however, present  some inherent and unavoidable features distinct with respect to their matter wave counterparts, like dissipation (due to radiation, or material absorption), and nonlinearity. Analysis, presented here, on the interplay between global dissipation, nonlinearity, and nonnormal dynamics not only offers an unconventional route to faithful information or energy transmission even in a dissipative environment, but also shed light on the possibility of anomalous single-mode lasing in these systems.

\bigskip



\begin{thebibliography}{100}
\bibitem{note} The terminology of `spectral ordinate' is coined here to refer the largest imaginary part of a complex energy spectrum, in contrast  to `spectral abscissa' which refers to the largest real part. Same argument is valid for the nomenclature of `numerical ordinate'.
\end{thebibliography}
\end{document}